\documentclass[slac_one]{revtex4}
\usepackage{graphicx}
\usepackage{fancyhdr}
\def\sigtot{$\sigma_{tot}$}
\begin{document}

\title{Zero momentum gluons  and the total  $pp$ and $p\bar{p}$  cross-section}

%

\author{A. Achilli and Y.N. Srivastava}
\affiliation{INFN and Physics Department University of Perugia, I-06123 Perugia, Italy }
\author{R.M. Godbole}
\affiliation{Centre for High Energy Physics, Indian Institute of Science,
 Bangalore, 560012, India}
\author{A. Grau}
\affiliation{ Departamento de F\'\i sica Te\'orica y del Cosmos, 
Universidad de Granada, 18071 Granada, Spain}
\author{G. Pancheri}
\affiliation{NFN Frascati National Laboratories,
Via Enrico Fermi 40, I-00044 Frascati, Italy}
\begin{abstract}
In this contribution we describe a QCD motivated model for total cross-sections which uses the eikonal representation and incorporates QCD mini-jets to drive the rise with energy of  \sigtot , while the impact parameter distribution is obtained through the Fourier transform of the $k_t$-distribution of soft gluons emitted in the parton-parton interactions giving rise to mini-jets in the final state. Using a phenomenological ans\"atz for the behaviour of the strong coupling constant $\alpha_s$ in the infrared region, our model gives a description of the total cross-section in agreement within limits imposed by the Martin-Froissart bound.
\end{abstract}

\maketitle

\thispagestyle{fancy}


\section{INTRODUCTION} 
We present here  our most recent results for a QCD motivated description of the total cross-section \cite{lastPLB}. The QCD contribution to the rise in the total cross-section had been advocated \cite{halzen} soon after the ISR results indicated that the total cross-sections  were increasing with energy by about 10\% above their previous values obtained in fixed target experiments. Subsequently, in order to  incorporate the QCD contribution as well as to satisfy  unitarity,   use was made of the eikonal representation, with the impact parameter distribution obtained from the hadronic form factors \cite{durand}. Although the rise of the total cross-section was correctly identified as being due to the rising number of low-x gluon collisions, these models  were usually unable to describe the gentle rise of the cross-section, and were not able to really connect with the rich jet phenomenology and the description of parton-parton scattering. Our aim has been to cure both of the above problems. First we have concentrated on calculating the so called mini-jet cross-sections through most commonly used DGLAP evoluted parton densities\cite{densities}, and then we have added soft gluon emission from initial state valence quarks to introduce acollinearity in parton-parton scattering.  Soft gluon emission in $k_t$ -space is naturally energy dependent, thus curing one of the problems of the description in terms of hadronic form factors. The other important difference with existing models, lies in our treatment of the infrared region. The model in fact takes its name from the Bloch-Nordsiek  (BN) theorem in QED, and its requirement of emission of an indefinite   number of soft quanta. When the QED re-summation procedure is transported to QCD, one encounters the problem of the infrared behaviour of the running coupling constant $\alpha_s$.  We believe that the contribution from the infrared region is very important for  the treatment of the total cross-section and have built our BN model on  the behaviour of $\alpha_s$ in the infrared. To do so, we have made use of an ans\"atz for such behaviour put forward long time ago to describe the intrinsic transverse momentum of Drell-Yan pairs \cite{oldkt}. Such ans\"atz  relates the rise with energy of the cross-section to the parameter regulating  the emission of zero momentum gluons.

\section{THE BLOCH-NORDSIECK MODEL FOR TOTAL CROSS-SECTIONS}
The usual impact parameter representation of a purely absorptive cross-section is written as
\begin{equation}
\sigma_{tot}(s) =
2\int [d^2 {\vec b}] [1-e^{-n(b,s)/2}] ,
\label{sigtot}
\end{equation}
where the eikonal function $n(b,s)$ is physically interpreted as the average number of collisions at a given impact parameter $b$ and squared c.m. energy  $s$.  We distinguish    between collisions calculable as QCD minijets, and  everything  else,  writing  the average number of collisions  as 
\begin{equation}
\label{nbs}
n(b,s)=n_{soft}(b,s) +n_{hard}(b,s) = n_{soft}(b,s) + A(b,s) \sigma_{jet}(s)
\end{equation} 
with  $n_{hard}$ including all outgoing parton processes with $p_t>p_{tmin}$. For its construction we use QCD minijets and a soft gluon distribution, which are described in the following two subsections along with a comparison with experimental data.

\subsection{QCD Mini-jets}
The mini-jet cross-section which drives the rise of the total cross-section is calculated in our model through the usual expression for inclusive jet production, namely
\begin{equation}
\sigma^{AB}_{\rm jet} (s,p_{tmin})= 
\int_{p_{tmin}}^{\sqrt{s}/2} d p_t \int_{4
p_t^2/s}^1 d x_1  \int_{4 p_t^2/(x_1 s)}^1 d x_2 
\sum_{i,j,k,l}
f_{i|A}(x_1,p_t^2) f_{j|B}(x_2,p_t^2)
  \frac { d \hat{\sigma}_{ij}^{ kl}(\hat{s})} {d p_t}.
  \end{equation}
where $A$ and $B$ are the colliding hadrons or photons, in this case $A-proton, B-proton/{\bar{p}}$. By construction,  this cross-section  depends on the particular parametrization of the DGLAP \cite{densities}
evoluted parton densities, some of which do extend to very low x-values but not too high $p_t^2$ values. See  \cite{lastPLB,ourlast,PRD60} for more details. 
This cross-section strongly depends on the lowest $p_t$ value upon which one integrates. The term {\it mini-jet} was introduced long ago \cite{jacob,corsetti,ddt,pp,collins} to indicate all those low $p_t$ processes  for which one can 
still expect them to be QCD calculable but which are actually not observed as hard jets.  $p_t$ being the scale at which to evaluate $\alpha_s$ in the mini-jet cross-section calculation, one can have  $p_{tmin}\approx 1\div 2\ GeV$.

\subsection{The Impact Parameter Distribution and Results}
To describe the distribution in $b-$space of the partons in the hadrons, which enter the eikonal representation, we  use the Fourier transform of the re-summed $k_t$ distribution of all soft gluons emitted in the parton-parton scattering, namely 
\begin{equation}\label{eq:abn}
 A_{BN}(b,s) =
{\cal N} \int d^2 {\bf K}_{\perp} {{d^2P({\bf K}_\perp)}\over{d^2 {\bf K}_\perp}} 
 e^{-i{\bf K}_\perp\cdot {\bf b}} \nonumber \\
 = {{e^{-h( b,q_{max})}}\over
 {\int d^2{\bf b} e^{-h(b,q_{max})} }}\equiv A_{BN}(b,q_{max}(s)).
 \label{Eq:abn}
\end{equation}
with 
\begin{equation}
h( b,q_{max}(s))  = 
\frac{16}{3}\int_0^{q_{max}(s) }
{{dk_t}\over{k_t}} 
 {{ \alpha_s(k_t^2) }\over{\pi}}   \left(\log{{2q_{\max}(s)}\over{k_t}}\right)\left[1-J_0(k_tb)\right] ,
\label{hdb}
\end{equation}  
In Eq. (\ref{hdb}), the upper limit of integration depends on the kinematics of the sub-process
\begin{equation}
parton_1+parton_2\rightarrow gluon+X
\end{equation}
while the lower limit of intergration is taken to be zero with
\begin{equation}
\alpha_s(k_t)= constant \times \left({{\Lambda}\over{k_t}}\right)^{2p}\ \ \ \ k_t\to 0
\label{phenoas}
\end{equation}
To relate this behaviour, inspired by the Richardson potential for charmonium, to the usual asymptotic freedom expression for $\alpha_s$, we have used the phenomenological expression
\begin{equation}
\alpha_s(k_t^2)={{12 \pi}\over{33-2N_f}} {{p}\over{\log[1+p({{ k_t^2}\over{\Lambda^2 }})^p]
 }}
\end{equation}
It is clear that the closer $p$ is to 1, the bigger the soft gluon integral $h(b,q_{max}(s))$ is and the stronger the saturation effects will be.
 
The final results for the $pp$ and $p\bar{p}$ total cross-sections and their comparison
with data and some other models are shown in Fig.(1).
\begin{figure*}[t]
\centering
\includegraphics[width=135mm]{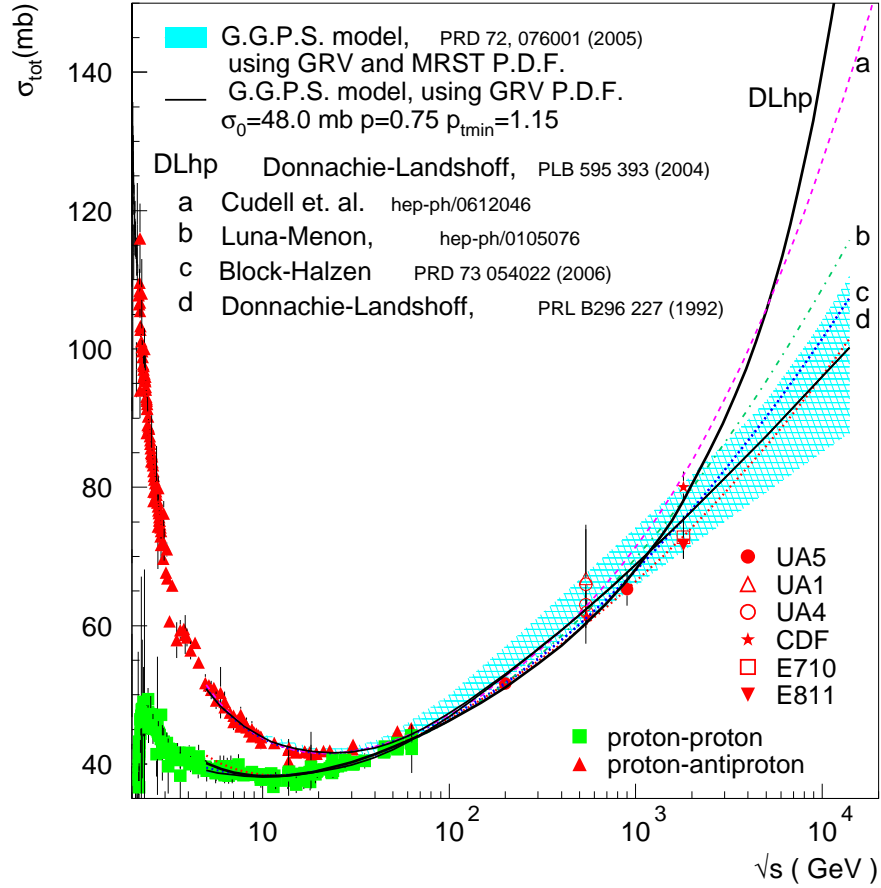}
\caption{Experimental data on proton-proton and proton-antiproton  total cross-sections\cite{data}
and comparison with our BN model described in the text. For a  discussion of
comparison with other models and explanation of symbols see
\protect\cite{lastPLB}.} \label{Fig:TX}
\end{figure*}

\begin{acknowledgments}
G.P. wishes to thank the Boston University Theoretical Physics group for hospitality while this contribution was prepared. This work has been partially supported by MEC (FPA2006-05294) and  Junta de Andaluc\'\i a (FQM 101 and FQM 437) and by INFN, Italy.
\end{acknowledgments}

\end{document}